\documentclass[%
superscriptaddress,
preprint,
 amsmath,amssymb,
prb,
]{revtex4-2}

\usepackage{graphicx}% Include figure files
\usepackage{dcolumn}% Align table columns on decimal point
\usepackage{bm}% bold math
\usepackage{color}
\usepackage{mathptmx}

\begin{document}

\title{Observation of terahertz magnon of Kaplan-Kittel exchange resonance in yttrium-iron garnet by Raman spectroscopy}% Force line breaks with \\

\author{Wei-Hung Hsu}
\email{jacky81418@gmail.com}
\affiliation{Department of Physics, Kyushu University, Fukuoka 819-0395, Japan}
\author{Ka Shen}
\affiliation{Center for Advanced Quantum Studies and Department of Physics, Beijing Normal University, Beijing 100875, China}
\author{Yasuhiro Fujii}
\affiliation{Department of Physical Sciences, Ritsumeikan University, Kusatsu 525-8577, Japan}
\author{Akitoshi Koreeda}
\affiliation{Department of Physical Sciences, Ritsumeikan University, Kusatsu 525-8577, Japan}
\author{Takuya Satoh}
\email{satoh@phys.titech.ac.jp}
\affiliation{Department of Physics, Kyushu University, Fukuoka 819-0395, Japan}
\affiliation{Department of Physics, Tokyo Institute of Technology, Tokyo 152-8551, Japan}

\date{\today}% It is always \today, today,
  % but any date may be explicitly specified

\begin{abstract}
Backscattering Raman spectroscopic investigations were performed on an yttrium-iron garnet single crystal using linearly and circularly polarized light. A terahertz (THz) magnon of the Kaplan-Kittel (KK) exchange resonance was discovered, which had been regarded as unobservable via optical methods. The KK exchange resonance had a 7.8-THz frequency at 80 K, and the polarization selection rule led to an antisymmetric Raman tensor of the $A_{2}$ mode. Moreover, the assignment of all the Raman-active phonon modes, 3$A_{\rm 1g}$, 8$E_{\rm g}$, and 14$T_{\rm 2g}$, was proposed. This study will stimulate further investigation of the coupling of THz magnons and phonons and pave the way toward THz optomagnonics.

\end{abstract}

\maketitle

\section{Introduction}

Yttrium-iron garnet ($\rm Y_{3}Fe_{5}O_{12}$, YIG), which possesses superior properties such as long-lived spin waves with very low damping ($\alpha \approx 10^{-5}$) and a high Curie temperature ($T\rm{_{c} = 560~K}$), has been regarded as an important magnetic material since its discovery over 60 years ago~\cite{Bertaut1956,Geller1957-2,Gilleo1986,Cherepanov1993}. YIG is a ferrimagnetic insulator that is one of the best candidates for applications in spintronics~\cite{Chumak2015}, magnonics~\cite{Kruglyak2010,Tabuchi2014}, and spin caloritronics~\cite{Bauer2012}. For developing magnetic devices with ultrafast responses, studies on terahertz (THz) magnon mechanisms are inevitable~
\cite{Satoh2010,Ivanov2014,Bossini2017,Baltz2018,Nemec2018,Kimel2020}.
However, previous studies on YIG mostly focused on applications of acoustic magnons with frequencies in the gigahertz (GHz) range~\cite{Dillon1957,LeCraw1958,Auld1967,Hu1971,Sandercock1973}. Existing studies provide limited information about the properties of THz optical magnons in YIG ~\cite{Cherepanov1993}. Recent investigations of full magnon band structures have been based on several theoretical methods~\cite{Barker2016,Xie2017,Liu2017,Shen2018,Barker2019}.
In experimental studies, the THz magnons in YIG have been identified only using inelastic neutron scattering spectroscopy~\cite{Plant1977,Princep2017,Shamoto2018,Nambu2020}. Hence, to understand the behaviors of the THz magnons in YIG in further detail, it is necessary to investigate other optical properties such as the selection rule using Raman spectroscopy.

YIG contains 20 Fe atoms in the primitive bcc unit cell. The magnetic moment is carried by Fe$^{3+}$ spins in the 12 tetrahedral $d$-sites (majority) and 8 octahedral $a$-sites (minority) with an antiparallel state in the unit cell, forming a two-sublattice ferrimagnet~\cite{Gilleo1986}. Among the 20 magnon modes, there are two types of modes with opposite precession directions.
Twelve modes belong to the counterclockwise (along the direction of the applied magnetic field) mode, which has a relatively low frequency and inconspicuous temperature-dependent frequency shift.
The other eight modes belong to the clockwise mode, which has a relatively higher frequency and significant temperature-dependent frequency shift~\cite{Barker2016}.
On the other hand, in a ferrimagnet with two sublattices, there are two uniform modes at wave number $k = 0$, namely, an acoustic mode (Kittel mode) with counterclockwise rotation and an optical mode with clockwise rotation~\cite{Douglass1960,Harris1963,Brinkman1966_1}. The optical mode is identical to the Kaplan-Kittel (KK) exchange resonance~\cite{Kaplan1953} between $a$-site and $d$-site Fe sublattices, which has been considered unobservable via optical [infrared (IR) or Raman] methods~\cite{Douglass1960} and has not been observed to date~\cite{Haussler1981,Grunberg1971}.

Therefore, in this study, we investigated the magnetic and crystallographic properties of a single-crystal YIG using polarized-backscattering Raman spectroscopy. Considering the temperature dependence of Raman shift from 80 to 140 K, we clearly identified a THz magnon of the KK exchange resonance. With regard to the crystallographic properties, although the phonon modes of YIG have been studied using Raman spectroscopy for decades, not all the phonon modes have been observed~\cite{Grunberg1971,Song1973,Mallmann2013,Costantini2015}. Here, using linearly and circularly polarized light, we completed the phonon-mode assignment of YIG.

\section{Materials and Methods}

\begin{figure}[t]
\includegraphics[width=8.6cm]{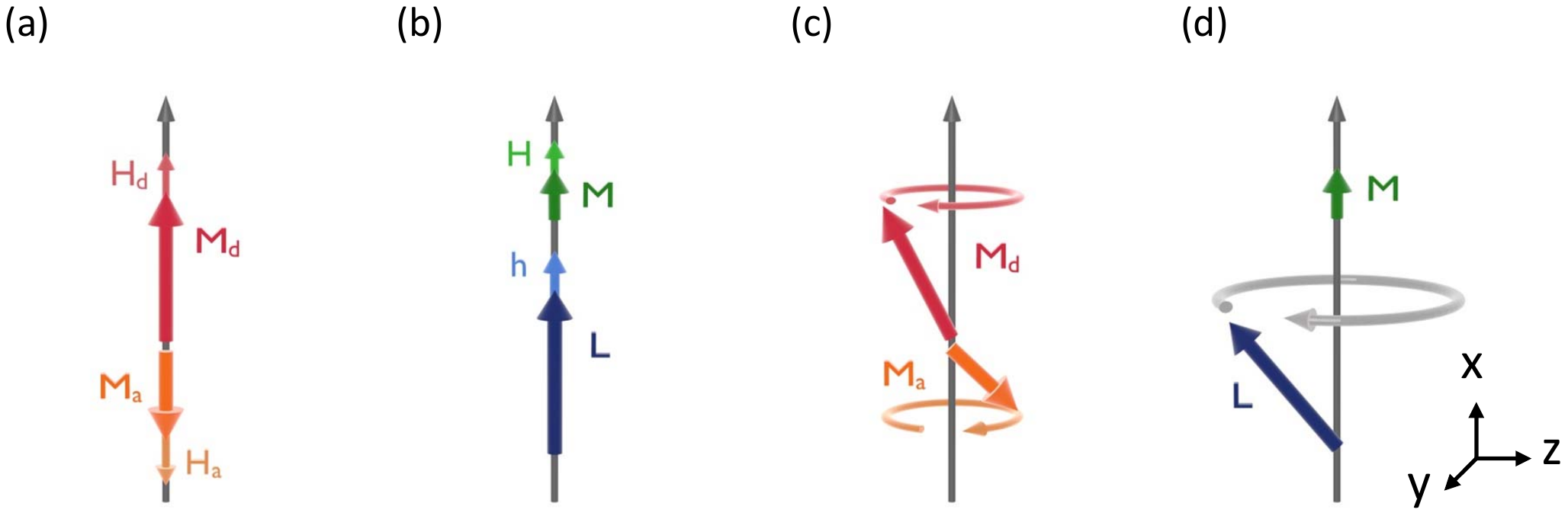}
\caption{Schematic of the KK exchange resonance mode.
(a) Sublattice magnetizations ${\bf M}_{\rm a}$ and ${\bf M}_{\rm d}$ under effective magnetic fields ${\bf H}_{\rm a}$ and ${\bf H}_{\rm d}$ in the ground state, respectively.
(b) Ferromagnetic and antiferromagnetic vectors ${\bf M}$ and ${\bf L}$ under effective magnetic fields ${\bf H}$ and ${\bf h}$ in the ground state, respectively.
(c) In the KK exchange resonance, ${\bf M}_{\rm a}$ and ${\bf M}_{\rm d}$ precess in the clockwise direction.
(d) ${\bf L}$ also precesses in the same direction, while ${\bf M}$ is stationary.} \label{fig_KKexchange}
\end{figure}

The $a$-site and $d$-site magnetizations of the Fe sublattices in YIG are defined as ${\bf M}_{\rm a}$ and ${\bf M}_{\rm d}$, respectively. Under an effective magnetic field, the magnetization during precessional motion in response to torques is described by the Landau-Lifshitz-Gilbert (LLG) equation. The LLG equations for ${\bf M}_{\rm a}$ and ${\bf M}_{\rm d}$ neglecting damping can be written as follows:
\begin{equation} \label{MagsubMo}
\begin{split}
\frac{d{\bf M}_{\rm a}}{dt} & = -\gamma {\bf M}_{\rm a} \times {\bf H}_{\rm a},\\
\frac{d{\bf M}_{\rm d}}{dt} & = -\gamma {\bf M}_{\rm d} \times {\bf H}_{\rm d},
\end{split}
\end{equation}
where $\gamma$ is the gyromagnetic ratio of iron that is invariant in both $a$- and $d$-site sublattices. The effective magnetic fields ${\bf H}_{\rm a}$ and ${\bf H}_{\rm d}$ acting on ${\bf M}_{\rm a}$ and ${\bf M}_{\rm d}$ can be expressed as ${\bf H}_{\rm a} = -\lambda{\bf M}_{\rm d}$ and ${\bf H}_{\rm d} = -\lambda{\bf M}_{\rm a}$, respectively, as shown in Fig.~\ref{fig_KKexchange}(a). The Weiss constant $\lambda$ is positive.
Assuming ${\bf M}_{\rm a} = \left ( -M_{\rm a},m^{y}_{\rm a},m^{z}_{\rm a} \right )$ and ${\bf M}_{\rm d} = \left ( M_{\rm d},m^{y}_{\rm d},m^{z}_{\rm d} \right )$, the precessional motions with time dependence ${\rm exp}(i\omega t)$ can be derived from Eq.~(\ref{MagsubMo}). The frequency of the KK exchange resonance is obtained as follows~\cite{Kaplan1953}:
\begin{equation} \label{KK10}
\begin{split}
\omega = \lambda\gamma \left | M_{\rm a} - M_{\rm d} \right |.
\end{split}
\end{equation}
Here, the effective fields arising from magnetic anisotropy, demagnetization, and magnetic dipoles are omitted because their contributions to the resonance frequency are negligibly small.

Because the magnetizations in the $a$- and $d$-site sublattices orientate in opposite directions with strength ratio $\left | {\bf M}_{\rm d} \right | : \left | {\bf M}_{\rm a} \right | = 3 : 2$, the ferromagnetic ${\bf M} = {\bf M}_{\rm a} + {\bf M}_{\rm d}$ and antiferromagnetic vectors ${\bf L} = {\bf M}_{\rm d} - {\bf M}_{\rm a}$ coexist in YIG [see Fig.~\ref{fig_KKexchange}(b)].
${\bf M}$ and ${\bf L}$ would also exhibit precessional motion under the effective fields ${\bf H}$ and ${\bf h}$, expressed as follows:
\begin{equation} \label{KK11}
\begin{split}
{\bf H} & = {\bf H}_{\rm d} + {\bf H}_{\rm a} = -\lambda {\bf M},\\
{\bf h} & = {\bf H}_{\rm d} - {\bf H}_{\rm a} = \lambda {\bf L}.
\end{split}
\end{equation}
The LLG equations for the vectors {\bf M} and {\bf L} are as follows:
\begin{equation} \label{KK12}
\begin{split}
\frac{d{\bf M}}{dt} &= -\frac{\gamma}{2} \left ( {\bf M}\times{\bf H} + {\bf L}\times{\bf h} \right ),\\
\frac{d{\bf L}}{dt} &= -\frac{\gamma}{2} \left ( {\bf M}\times{\bf h} + {\bf L}\times{\bf H} \right ).
\end{split}
\end{equation}
Assuming ${\bf M} = \left ( M,m_{y},m_{z} \right )$ and ${\bf L} = \left ( 5M,l_{y},l_{z} \right )$ according to the ratio of ${\bf M}_{\rm d}$ and ${\bf M}_{\rm a}$ in YIG, the time-dependent motions of ${\bf M}$ and ${\bf L}$ in the $\it y$-$\it z$ plane are expressed as follows:
\begin{equation} \label{KK13}
\begin{split}
\frac{dm_{y}}{dt} = \frac{dm_{z}}{dt} = 0,
\end{split}
\end{equation}
\begin{equation} \label{KK14}
\begin{split}
\frac{dl_{y}}{dt} &= -\lambda\gamma M \left ( 5m_{z} - l_{z} \right ),\\
\frac{dl_{z}}{dt} &= -\lambda\gamma M \left ( l_{y} - 5m_{y} \right ).
\end{split}
\end{equation}
Equation~(\ref{KK13}) shows that the ferromagnetic vector ${\bf M}$ is not in precessional motion [see Fig.~\ref{fig_KKexchange}(d)]; this implies that the projecting vectors of the $\it y$-$\it z$ plane, namely, ${\bf M}_{\rm d}$ and ${\bf M}_{\rm a}$, are not only antiparallel but also of the same magnitude [see Fig.~\ref{fig_KKexchange}(c)]. On the other hand, the antiferromagnetic vector ${\bf L}$ precesses clockwise, and the precessional frequency obtained using Eq.~(\ref{KK14}) is the same as that obtained using Eq.~(\ref{KK10}), namely, $\omega = \lambda\gamma M$.
This frequency is equivalent to $(12S_{\rm d}-8S_{\rm a}) \left | J_{\rm ad} \right | = 10 \left | J_{\rm ad} \right |$, where $S_{\rm a} = S_{\rm d} = 5/2$ and $J_{\rm ad}$ is the exchange constant between the magnetizations at the $a$- and $d$-site sublattices~\cite{Shen2018}. Thus, the KK exchange resonance is identical to the clockwise uniform mode at $k = 0$~\cite{Douglass1960,Harris1963} ($\alpha_1$ mode in Ref.~\cite{Shen2018}).

\begin{figure}[t]
\includegraphics[width=8.6cm]{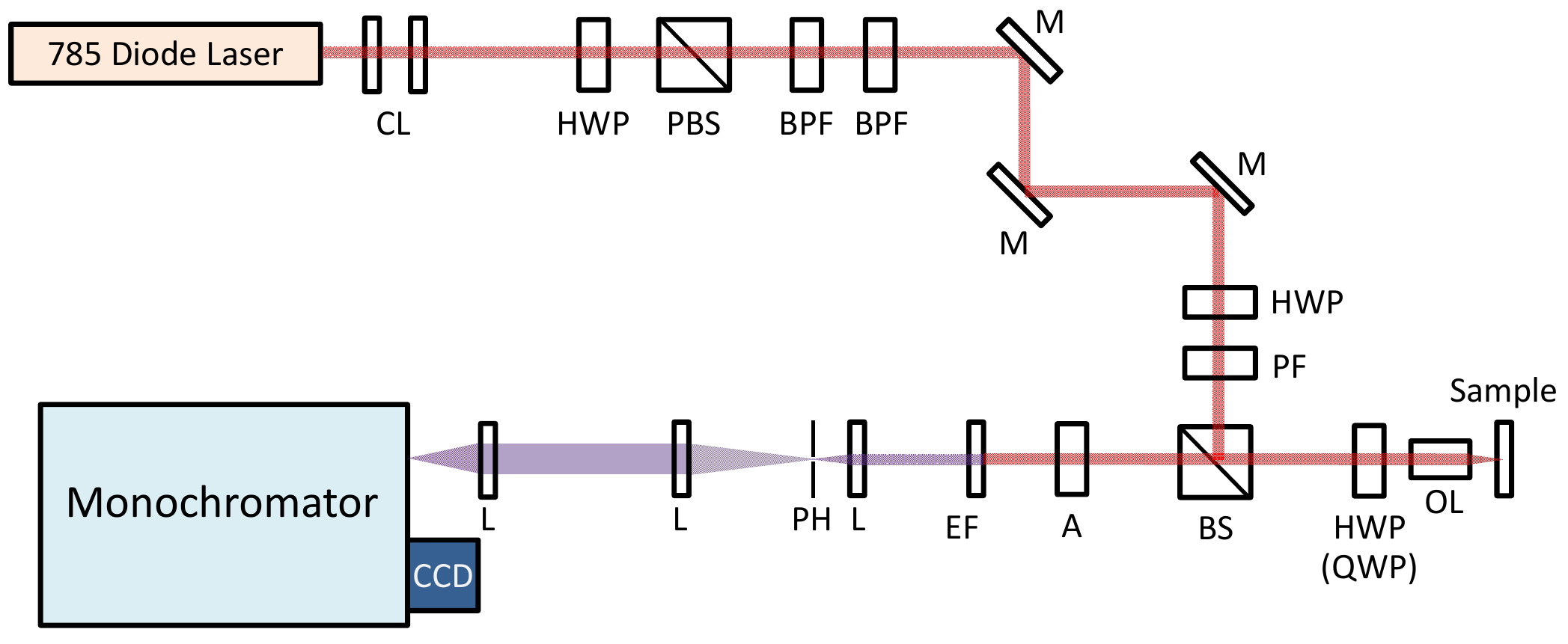}
\caption{Homemade backscattering Raman-spectroscopy system. The incident light from the 785-nm diode laser (LD785-SEV300, Thorlabs) passes through a collimating lens (CL), half-wave plate (HWP), polarized beam splitter (PBS), bandpass filters (BPF), polarized filter (PF), beam splitter (BS), quarter-wave plate (QWP), and objective lens (OL). The diameter of the focused laser spot equals 25 $\mu$m. After sample irradiation by the incident polarized light at 40 mW, the scattered light from the sample is polarized using a Glan--Taylor prism (denoted by A). Subsequently, the inelastically scattered light is separated from its elastically scattered counterpart using an edge filter (EF), combination of lenses (L), and pinhole (PH). The scattered light comprising different wavelengths is spatially separated by the grating (1200 g/mm grating blazed at 500 nm) in the monochromator (Acton SpectraPro 2500i, Princeton Instruments) followed by subsequent detection by CCD image sensors (S11501-1007S, Hamamatsu Photonics). The frequency resolution of this Raman system equals 1.0 $\rm cm^{-1}$. The heating caused by the focused laser spot results in a 1--2 $\rm cm^{-1}$ redshift in the Raman spectrum.}
\label{fig:Raman}
\end{figure}

\begin{figure}[t]
\centering
\includegraphics[width=8.6cm]{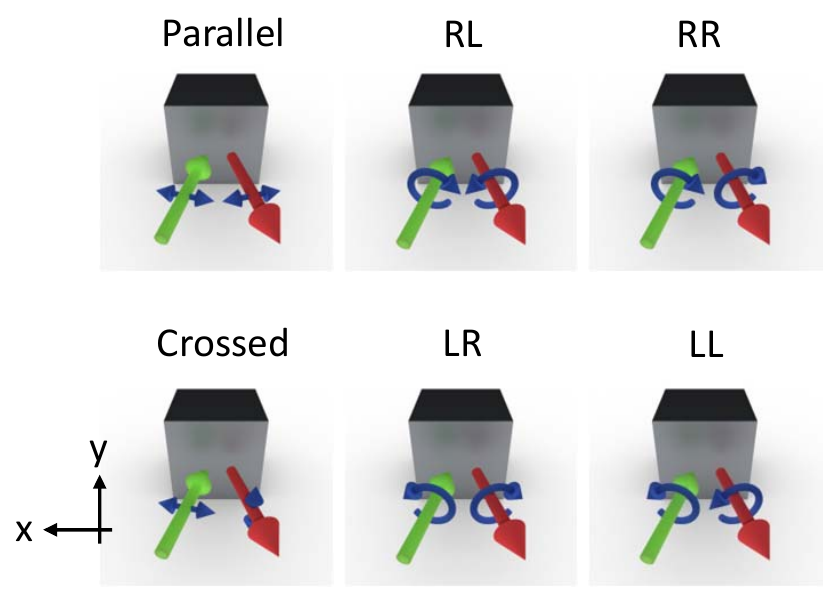}
\caption{Schematics of linearly and circularly polarized configurations for selection rules. The parallel- and crossed-polarized light rays indicate that the polarizations of the incident and selected scattered light are parallel and perpendicular, respectively. R and L represent the right- and left-handed helicities, respectively, of the circularly polarized light.}
\label{fig:Pol}
\end{figure}

In the proposed 180$^{\circ}$ backscattering Raman-spectroscopy system (Fig.~\ref{fig:Raman}), the diode-laser radiation with 785-nm wavelength was linearly and circularly polarized using polarizers as well as half- and quarter-wave plates. In accordance with the selection of polarization of the incident and scattered light, two linearly polarized configurations [parallel ($\parallel$) and crossed ($\perp$)] and four circularly polarized configurations (RL, LR, RR, and LL) were defined
as shown in Fig.~\ref{fig:Pol}. The helicity of the circularly polarized light was defined as the direction of polarization rotation in the sample plane, irrespective of the propagation direction.

The sample was mounted under vacuum maintained inside an optical cryostat (VPF-100, JANIS) capable of controlling temperatures over a wide range from the boiling point of liquid nitrogen to 500 K. The temperature sensor was attached to the sample holder made of copper.

Before we searched for the KK exchange resonance mode, we investigated the phonon modes using angle-resolved polarized Raman spectroscopy~\cite{Fujii2014} to distinguish the magnon and phonon modes.
A YIG single crystal grown by the floating zone technique was oriented along the [111] direction and polished to a thickness of 160 $\rm \mu m$. YIG belongs to the cubic centrosymmetric space group $Ia\bar{3}d$ and crystallographic point group $m\bar{3}m$~\cite{Geller1957-2}. The phonon Raman tensors are described in the Appendix, where the bases of the Raman tensors~\cite{Hayes1978} are transformed to ${\hat{\bf x}}\parallel [11\bar{2}]$, ${\hat{\bf y}}\parallel [\bar{1}10]$, and ${\hat{\bf z}}\parallel [111]$.

For the crystallographic point group $m\bar{3}m$, all three Raman-active phonon modes, namely, $A_{\rm 1g}$, $E_{\rm g}$, and $T_{\rm 2g}$, are not expected to have polarization azimuth dependence on linearly polarized light propagating along the $z$ direction. On the other hand, we have the Jones vectors $\left |{\rm H} \right \rangle=(1,0)$ for horizontal polarization, $\left |{\rm V} \right \rangle=(0,1)$ for vertical polarization, $\left |{\rm R} \right \rangle=(1,-i)/\sqrt{2}$ for right-handed circular polarization, and $\left |{\rm L} \right \rangle=(1,i)/\sqrt{2}$ for left-handed circular polarization~\cite{Yariv2006}. Therefore, the Raman intensity ratios [$I_{\rm \parallel} : I_{\rm \perp} : I_{\rm RL} : I_{\rm RR}$] of $A_{\rm 1g}$, $E_{\rm g}$, and $T_{\rm 2g}$ are calculated to be [$1 : 0 : 0 : 1$], [$1 : 1 : 2 : 0$], and [$3 : 2 : 4 : 1$], respectively~\cite{Khan2019}. Further, $I_{\rm LR} = I_{\rm RL}$, and $I_{\rm LL} = I_{\rm RR}$.
In addition, an in-plane magnetic field of 2.5 kOe was applied using permanent magnets to saturate our sample magnetically. This ensured that Faraday rotation of the incident and scattered light in the sample did not influence the Raman selection rules.

\section{Results and Discussion}

\begin{figure*}
\includegraphics[width=17.5cm]{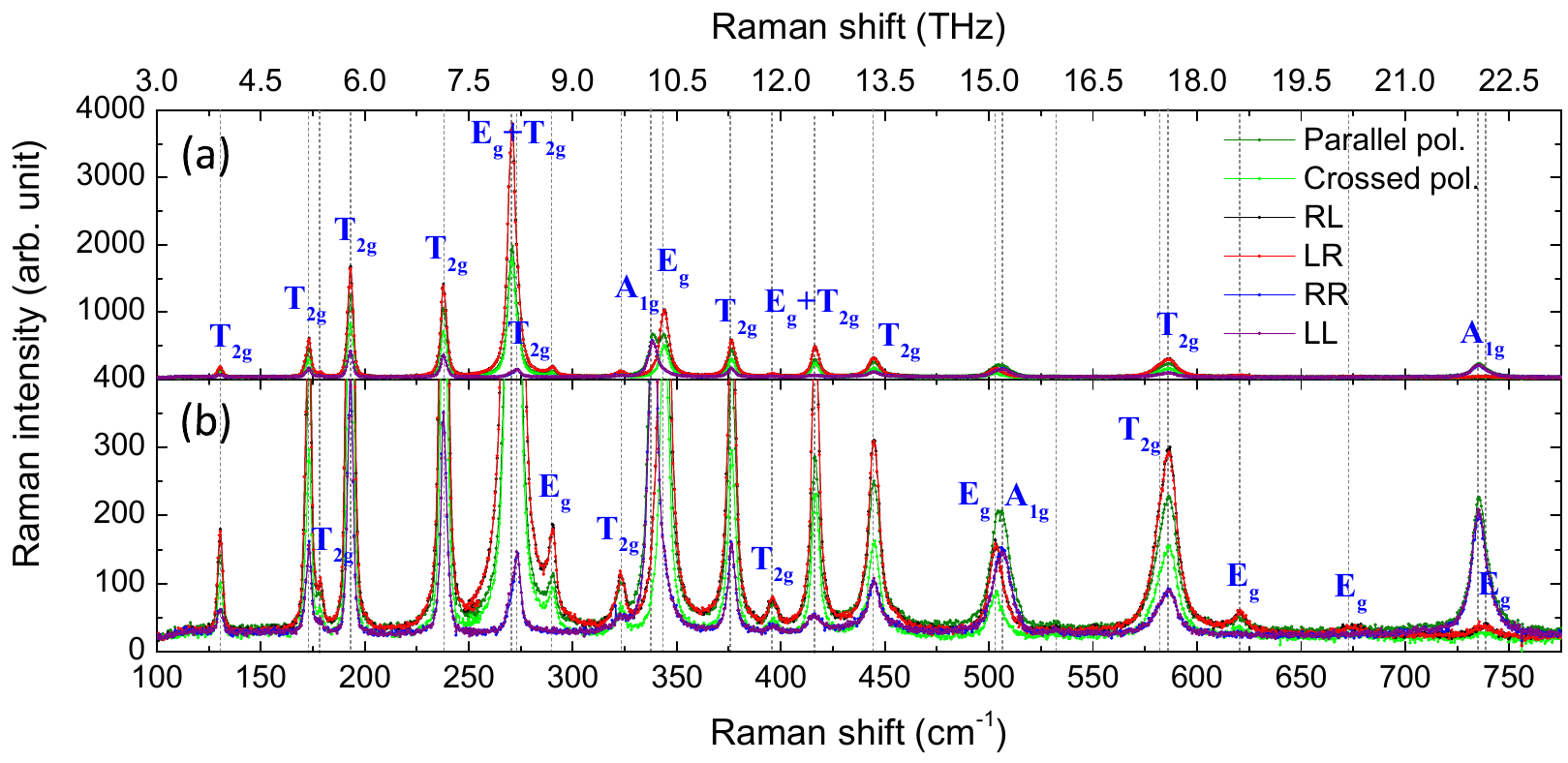}
\caption{(a) Full Raman spectra with various polarization configurations at 80 K. For linear polarization, dark and light green spectra represent parallel- and crossed-polarized configurations, respectively. For circular polarization, black, red, blue, and purple spectra represent the RL-, LR-, RR-, and LL-polarized configurations, respectively. (b) Magnification of (a).}
\label{fig_Fullspec}
\end{figure*}

Figures~\ref{fig_Fullspec}(a) and \ref{fig_Fullspec}(b) show the Raman spectra ranging from 100 to 775 cm$^{-1}$ at 80 K and their magnifications, respectively. In previous studies, with only linearly polarized configurations, some ambiguous peaks could not be distinguished~\cite{Grunberg1971,Song1973}. However, in the present study, using both linearly and circularly polarized configurations, all the phonon modes, namely, 3$A_{\rm 1g}$, 8$E_{\rm g}$, and 14$T_{\rm 2g}$, in YIG were successfully assigned and are listed in Table~\ref{tab_assignment}.

\begin{table}[b]
\caption{\label{tab:Ass}%
Assignment of phonon signals in YIG $(111)$ Raman spectra at 80~K. All shifts are in $\rm{cm^{-1}}$.
}
\begin{ruledtabular}
 \begin{tabular}{cccccc}
Ref. \cite{Grunberg1971} & Mode & Ref. \cite{Song1973} & Mode & This study & Mode \\
 \colrule
130 & $T_{\rm 2g}$ & 131 & $T_{\rm 2g}$ & 132 & $T_{\rm 2g}$ \\
175 & $T_{\rm 2g}$ & 174 & $T_{\rm 2g}$ & 174 & $T_{\rm 2g}$ \\
 & & & & 178 & $T_{\rm 2g}$ \\
193 & $T_{\rm 2g}$ & 194 & $T_{\rm 2g}$ & 194 & $T_{\rm 2g}$ \\
237 & $T_{\rm 2g}$ & 238 & $T_{\rm 2g}$ & 239 & $T_{\rm 2g}$ \\
274 & $E_{\rm g}$ & 274 & $E_{\rm g} + T_{2g}$ & 272 & $E_{\rm g} + T_{2g}$ \\
 & & & & 274 & $T_{\rm 2g}$ \\
 & & & & 290 & $E_{\rm g}$ \\
315 & $E_{\rm g}$ & 319 & $E_{\rm g}$ & & \\
 & & 324 & $T_{\rm 2g}$ & 323 & $T_{\rm 2g}$ \\
 & & & & 340 & $A_{\rm 1g}$ \\
347 & $E_{\rm g} + A_{1g}$ & 346 & $E_{\rm g}$ & 345 & $E_{\rm g}$ \\
380 & $T_{\rm 2g}$ & 378 & $T_{\rm 2g}$ & 377 & $T_{\rm 2g}$ \\
 & & & & 396 & $T_{\rm 2g}$ \\
 & & 416 & $E_{\rm g}$ & 416 & $E_{\rm g} + T_{2g}$ \\
420 & $E_{\rm g} + T_{2g}$ & 419 & $T_{\rm 2g}$ & & \\
449 & $T_{\rm 2g} + A_{1g}$ & 445 & $T_{\rm 2g}$ & 446 & $T_{\rm 2g}$ \\
 & & 456 & $E_{\rm g}$ & & \\
 & & & & 505 & $E_{\rm g}$ \\
507 & $A_{\rm 1g}$ & 504 & $A_{\rm 1g}$ & 507 & $A_{\rm 1g}$ \\
 & & & & 581 & $T_{\rm 2g}$ \\
593 & $T_{\rm 2g}$ & 592 & $T_{\rm 2g}$ & 587 & $T_{\rm 2g}$ \\
 & & 624 & $E_{\rm g}$ & 620 & $E_{\rm g}$ \\
 & & & & 675 & $E_{\rm g}$ \\
 & & 685 & $E_{\rm g} + T_{2g}$ & & \\
698 & $A_{\rm 1g} + E_{g}$ & 692 & $E_{\rm g} + T_{2g}$ & & \\
 & & 704 & $A_{\rm 1g}$ & & \\
 & & 711 & $T_{\rm 2g}$ & & \\
 & & 736 & $A_{\rm 1g}$ or & 737 & $A_{\rm 1g}$ \\
740 & $A_{\rm 1g}$ & & $E_{\rm g} + T_{2g}$ & 739 & $E_{\rm g}$ \\
 \end{tabular}
\label{tab_assignment}
\end{ruledtabular}
\end{table}

\begin{figure}[t]
\includegraphics[width=8.6 cm]{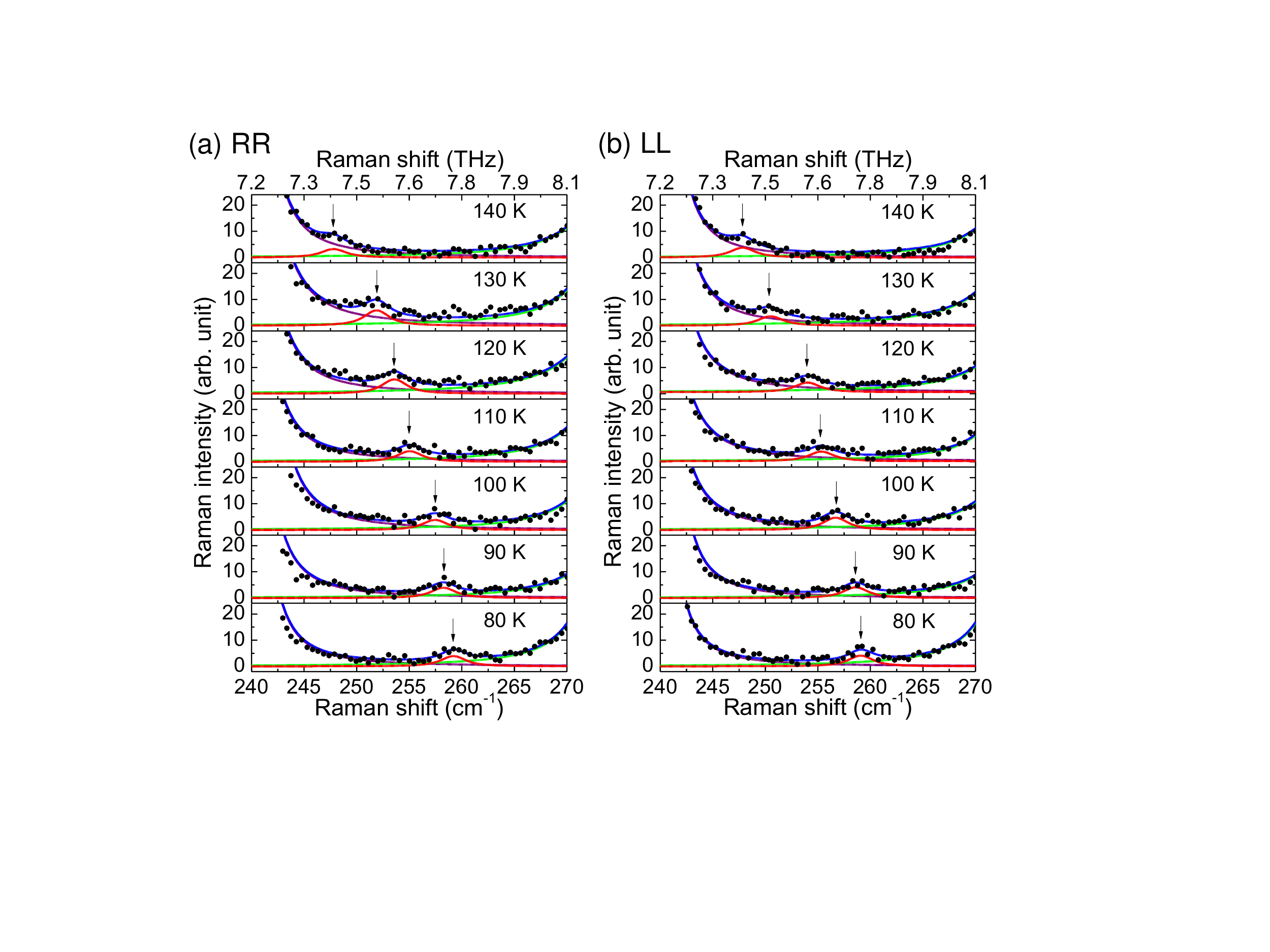}
\caption{YIG (111) Raman spectra in the 240--270-$\rm cm^{-1}$ range with (a) RR- and (b) LL-polarized configurations in the 80--140 K temperature range. The red, green, and purple curves fit the magnon and two $T_{\rm 2g}$ phonons, respectively. The blue curve denotes the sum of the red, green, and purple curves. Fitting was performed using a Voigt function that represents the convolution of the Gaussian and Lorentzian functions with full widths at the half maxima of 1.0 and 2.5 $\rm cm^{-1}$, respectively. Unlike other phonon signals, the magnon peaks indicated by arrows represent significant temperature-dependent frequency shifts.}
\label{fig_MagTemp}
\end{figure}

By considering the THz magnon frequency range reported in~\cite{Xie2017,Liu2017,Princep2017,Shen2018,Shamoto2018,Barker2019}, we focused on the spectra in the 240--270-$\rm cm^{-1}$ range, as depicted in Figs.~\ref{fig_MagTemp}(a) and \ref{fig_MagTemp}(b) for the RR- and LL-polarized configurations, respectively. For the RR-polarized configuration, we observed a tiny signal to shift significantly from 80 to 140 K and blend into the tail of the $T_{\rm 2g}$ phonon signal above 140 K. Additionally, a nearly identical signal was observed for the LL-polarized configuration. Because of the significant temperature dependence of the frequency shift, the tiny signal was assigned to the THz magnon. The wave number $k$ of the detected magnon in the 180$^{\circ}$ backscattering geometry was 4$\pi n/\lambda$ ($n$ is the refractive index, $\lambda$ is the incident light wavelength). Therefore, $a_{0} k\approx 0.05~{\rm rad} \ll 1$ (the lattice constant of YIG $a_{0} = 1.24$ nm~\cite{Geller1957-2}), meaning that the THz magnon of the KK exchange resonance with $k \approx 0$ was excited at 7.8 THz, the frequency corresponding to 260 $\rm cm^{-1}$ at 80 K.

\begin{figure}[t]
\includegraphics[width=8.6 cm]{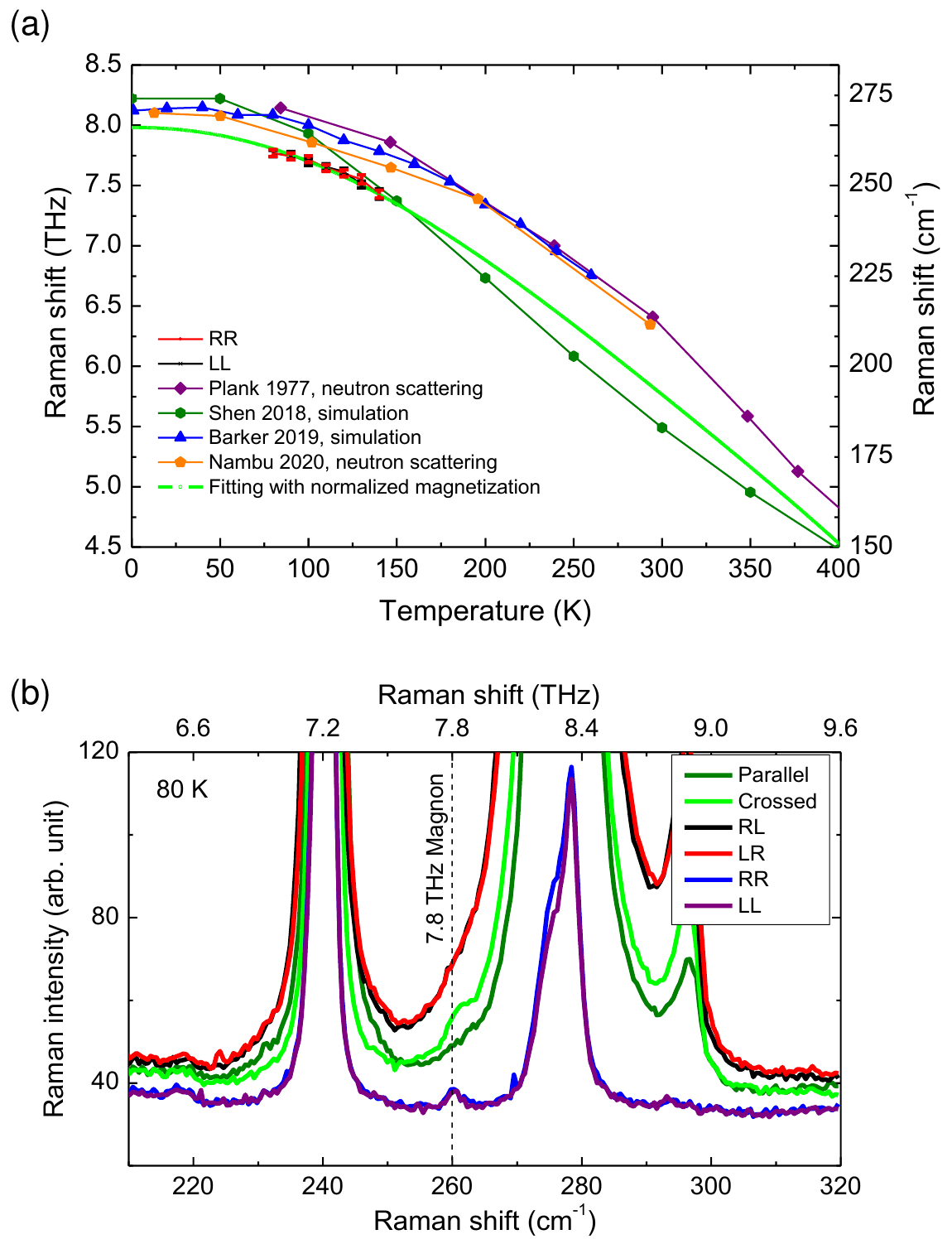}
\caption{(a) Temperature-dependent frequency shift in the KK mode. The red and black lines with error bars denote the KK mode observed in our Raman spectra with RR- and LL-polarized configurations, respectively. The purple line with squares ~\cite{Plant1977} and orange line with pentagons ~\cite{Nambu2020} represent the experimental data for neutron scattering.
The blue line with triangles ~\cite{Barker2019} and green line with hexagons ~\cite{Shen2018} indicate the simulation results obtained using different models. The KK mode is fitted with normalized magnetization (yellowish green)~\cite{Anderson1964}.
(b) The selection rule for the KK mode of YIG observed with crossed-, RR-, and LL-polarized configurations at 7.8 THz and 80 K.}
\label{fig_MagSele}
\end{figure}

Furthermore, the temperature-dependent frequency shifts of the THz magnon in our RR and LL spectra were identified and compared to the results obtained from neutron scattering measurements ~\cite{Plant1977,Nambu2020} and simulations~\cite{Shen2018,Barker2019}, as shown in Fig.~\ref{fig_MagSele}(a). Our results show the same tendency as reported in other studies. However, there is a frequency deviation between our results and the others. The exchange constant $J_{\rm ad}$ was determined in previous studies based on the results of magnetization, specific heat measurement, inelastic neutron scattering, or first-principles calculations.
However, we consider that Raman spectroscopy provides a more precise frequency resolution than other techniques.
By extrapolating the magnon frequency of 7.8 THz at 80 K down to 4 K based on the temperature dependence of magnetization~ \cite{Anderson1964}, we obtained a frequency of 8.0 THz at 4 K. This yields $J_{\rm ad} = -38$ K according to $10 \left | J_{\rm ad} \right | = 8.0~{\rm THz}$.
Furthermore, the in-plane magnetic field of 2.5 kOe shifted the magnon frequency by a few GHz, which is well below the experimental resolution.

Moreover, we focused on the same frequency range and repeated the Raman measurements with all linearly and circularly polarized configurations at 80~K; the results are shown in Fig.~\ref{fig_MagSele}(b).
The Raman intensity ratio of the KK mode was [$I_{\rm \parallel} : I_{\rm \perp} : I_{\rm RL} : I_{\rm RR}] = [0 : 1 : 0 : 1]$, indicating that the magnon Raman tensor can be expressed in the antisymmetric form as
\[
{R}
=
\begin{pmatrix}
  0 & iK \\
  -iK & 0
\end{pmatrix}.
\]
This antisymmetric form of the KK mode corresponds to linear (first-order) magnetic excitation, where $K$ is proportional to $m_{ z}$~\cite{Fleury1968,Gall1971,Wettling1975,Borovik1982,Hisatomi2019}. The similarity of the KK mode to the one in the GHz acoustic magnon (Kittel mode) excited through light scattering~\cite{Auld1967,Hu1971,Sandercock1973} indicates that the KK and Kittel modes have the same symmetry.
In the KK mode, all the atomic spins from the same sublattice simultaneously rotate with their spins always parallel and with the same precession amplitude, similar to the Kittel mode in the ferromagnetic subsystem. The other sublattice behaves in the same way. Therefore, the direction of precessional motion of each sublattice is reversed under mirror operations ($\sigma_v$ or $\sigma_d$), which corresponds to the $A_{2}$ mode \cite{Cracknell1969}.
Note that the antisymmetric form of the Raman tensor and the corresponding $A_{2}$ mode cannot be assigned to the phonon mode under the $m\bar{3}m$ point group.

Symmetric contributions owing to quadratic (second order) excitation~
\cite{Dillon1970,Pisarev1971,Wettling1975,Ferre1984,Hisatomi2019}
were not observed in the present study, probably because the linear magneto-optical (Faraday) effect dominated the quadratic magneto-optical (Cotton-Mouton) effect at a wavelength of 785 nm~\cite{Wettling1975_1}.

In addition, unlike the acoustic magnon, the ferromagnetic vector $\bf M$ of the KK mode does not precess, as shown in Fig.~\ref{fig_KKexchange}(d). 
Therefore, the KK mode was considered to be unobservable via optical methods~\cite{Douglass1960}.
Nevertheless, the KK mode was observed in our Raman method as a linear magnetic excitation, which usually results from the precession of $\bf M$. This contradiction can be explained as follows.
The absorption coefficient of YIG at 785 nm is 50--100 $\rm cm^{-1}$~\cite{Dillon1959,Wood1967,Scott1974}.
This is attributed to off-resonant transition with the nearest resonances at 900 nm (${}^{6}A_{\rm 1g} \rightarrow {}^{4}T_{\rm 1g}$) and 700 nm (${}^{6}A_{\rm 1g} \rightarrow {}^{4}T_{\rm 2g}$) in octahedral Fe$^{3+}$ ions~\cite{Wood1967}.
The Faraday rotation of 600$^{\circ}$--800$^{\circ}$/cm~ at 785 nm \cite{Clogston1960,Takeuchi1973,Wettling1975_1,Scott1975,Hansen1983} dominantly originates from optical transitions in octahedral Fe$^{3+}$ ions~\cite{Pisarev1977,Hansen1983,Deb2012,Maehrlein2018}.
Therefore, the Faraday rotation is more sensitive to ${\bf M}_{\rm a}$ than ${\bf M}_{\rm d}$. The KK mode resulted from the precessional motion of ${\bf M}_{\rm a}$~\cite{Reid2010}.

The KK exchange resonance between iron and rare-earth sublattices is identified in rare-earth iron garnets through IR~\cite{Sievers1963,Yamamoto1974,Kang2010} and pump-probe~\cite{Parchenko2013,Parchenko2016,Chekhov2018} spectroscopy. However, in YIG, because the ferromagnetic vector ${\bf M}$ of the KK mode does not precess owing to the equality of $\gamma$ for $\bf{M}_{\rm a}$ and $\bf{M}_{\rm d}$, this mode cannot be observed by IR spectroscopy and magnetic resonance spectroscopy.

\section{Conclusions}
All the phonon modes, namely, 3$A_{\rm 1g}$, 8$E_{\rm g}$, and 14$T_{\rm 2g}$, in YIG were identified and assigned using linearly and circularly polarized light through Raman spectroscopy. A THz magnon of the KK exchange resonance in YIG was discovered in Raman spectra at temperatures from 80 to 140 K, and the exchange constant was $J_{\rm ad}$ = $-38$ K. The selection rule of magneto-optical coupling in YIG was experimentally confirmed, which suggested an antisymmetric magnon Raman tensor ($A_{2}$) corresponding to the linear magneto-optical effect.
This study will stimulate further investigation of the coupling of THz magnons and phonons for applications involving spin caloritronics and pave the way toward THz optomagnonics.

\section*{Acknowledgments}
We would like to thank J. Barker and G. E. W. Bauer for valuable discussions and a critical proofreading of the manuscript. We also thank
P. Maldonado, P. M. Oppeneer, L.-W. Wang, and K. Xia for sharing with us their data on phonon dispersion calculations.
This study was supported by the Japan Society for the Promotion of Science (JSPS) KAKENHI (Grants No. JP15H05454, No. JP17K18765, No. JP19H01828, No. JP19H05618, No. JP19K21854, and No. JP26103004) and the JSPS Core-to-Core Program (A. Advanced Research Networks).

\section*{Appendix: Phonon-mode Raman tensors}

The Raman tensors $R$ pertaining to the crystallographic point group $m\bar{3}m$ with ${\hat{\bf x}}\parallel [100]$, ${\hat{\bf y}}\parallel [010]$, and ${\hat{\bf z}}\parallel [001]$ can be expressed as follows~\cite{Hayes1978}:
\begin{equation}
R(A_{\rm 1g})
=
a
\begin{pmatrix} 
1 & 0 & 0 \\
0 & 1 & 0 \\
0 & 0 & 1 
\end{pmatrix},
\tag{A1}
\end{equation}
%
%\vspace{-5mm}
%
\begin{equation}
R(E_{\rm g}^1)
=
b
\begin{pmatrix} 
1 & 0 & 0 \\
0 & 1 & 0 \\
0 & 0 & -2
\end{pmatrix},
\nonumber \\
\end{equation}
%
%\vspace{-5mm}
%
\begin{equation}
R(E_{\rm g}^2)
=
b
\begin{pmatrix} 
-\sqrt{3} & 0 & 0 \\
0 & \sqrt{3} & 0 \\
0 & 0 & 0
\end{pmatrix},
\tag{A2}
\end{equation}
%
%\vspace{-5mm}
%
\begin{equation}
R(T_{\rm 2g}^{1})
=
d
\begin{pmatrix}
0 & 0 & 0 \\
0 & 0 & 1 \\
0 & 1 & 0
\end{pmatrix},
\nonumber \\
\end{equation}
%
%\vspace{-5mm}
%
\begin{equation}
R(T_{\rm 2g}^{2})
=
d
\begin{pmatrix} 
0 & 0 & 1 \\
0 & 0 & 0 \\
1 & 0 & 0
\end{pmatrix},
\nonumber \\
\end{equation}
%
%\vspace{-5mm}
%
\begin{equation}
R(T_{\rm 2g}^{3})
=
d
\begin{pmatrix} 
0 & 1 & 0 \\
1 & 0 & 0 \\
0 & 0 & 0
\end{pmatrix}.
\tag{A3}
\end{equation}\\[-10pt]
In the above equations, $a$, $b$, and $d$ indicate the nonzero Raman tensor elements.

Because our YIG sample was cut along the [111] direction, the original tensors were transformed from the [001] direction to the [111] direction using the transformation operator $T$ given by
\begin{equation} \label{Rot_oper}
\begin{split}
T =
\left ( \begin{matrix}
\frac{1}{\sqrt{6}} & -\frac{1}{\sqrt{2}} & \frac{1}{\sqrt{3}}\\ 
\frac{1}{\sqrt{6}} & \frac{1}{\sqrt{2}} & \frac{1}{\sqrt{3}}\\ 
-\frac{2}{\sqrt{6}} & 0 & \frac{1}{\sqrt{3}}
\end{matrix} \right ).
\end{split}
\tag{A4}
\end{equation}
This equation rotates ${\hat{\bf x}}\parallel [100]$, ${\hat{\bf y}}\parallel [010]$, and ${\hat{\bf z}}\parallel [001]$ to ${\hat{\bf x}}^{\prime}\parallel [11\bar{2}]$,  ${\hat{\bf y}}^{\prime}\parallel [\bar{1}10]$, and  ${\hat{\bf z}}^{\prime}\parallel [111]$, respectively. Subsequently, the Raman tensors $R^{\prime}={T}^{-1} R  {T}$ can be expressed as
\begin{equation}
R^{\prime}(A_{\rm 1g})
=
a
\begin{pmatrix}
    1 & 0 & 0 \\
    0 & 1 & 0 \\
    0 & 0 & 1
\end{pmatrix},
\tag{A5}
\end{equation}
%
%\vspace{-5mm}
%
\begin{equation}
R^{\prime}(E{_{\rm g}^1})
=
b
\begin{pmatrix}
    -1 & 0 &\sqrt{2}\\
    0 & 1 & 0  \\
    \sqrt{2} & 0 & 0
\end{pmatrix},
\nonumber \\
\end{equation}
%
%\vspace{-5mm}
%
\begin{equation}
R^{\prime}(E{_{\rm g}^2})
=
b
\begin{pmatrix}
    0 & 1 & 0 \\
    1 & 0 & \sqrt{2} \\
    0 & \sqrt{2} & 0
\end{pmatrix},
\tag{A6}
\end{equation}
%
%\vspace{-5mm}
%
\begin{equation}
R^{\prime}(T{_{\rm 2g}^1})
=
\frac{d}{6}
\begin{pmatrix}
    -4 & -2\sqrt{3} & -\sqrt{2}  \\
    -2\sqrt{3} & 0 & \sqrt{6} \\
    -\sqrt{2} & \sqrt{6} & 4
\end{pmatrix},
\nonumber \\
\end{equation}
%
%\vspace{-5mm}
%
\begin{equation}
R^{\prime}(T{_{\rm 2g}^2})
=
\frac{d}{6}
\begin{pmatrix}
    -4 & 2\sqrt{3} & -\sqrt{2}  \\
    2\sqrt{3} & 0 & -\sqrt{6} \\
    -\sqrt{2} & -\sqrt{6} & 4
\end{pmatrix},
\nonumber \\
\end{equation}
%
%\vspace{-5mm}
%
\begin{equation}
R^{\prime}(T{_{\rm 2g}^3})
=
\frac{d}{6}
\begin{pmatrix}
    2 & 0 & 2\sqrt{2}  \\
    0 & -6 & 0 \\
    2\sqrt{2} & 0 & 4
\end{pmatrix}.
\tag{A7}
\end{equation}\\[-10pt]

The Raman scattering intensities $I(\hat{\bf e}_{\rm i}, \hat{\bf e}_{\rm s})$ for $A_{\rm 1g}$, $E_{\rm g}$, and $T_{\rm 2g}$ can be expressed as
\begin{equation} \label{RamanIn111_A}
\begin{split}
I_{A_{\rm 1g}} (\hat{\bf e}_{\rm i}, \hat{\bf e}_{\rm s}) \propto \left | \hat{\bf e}_{\rm i}^{\ast} \cdot R^{\prime}(A_{\rm 1g}) \cdot \hat{\bf e}_{\rm s} \right |^{2},
\end{split}
\tag{A8}
\end{equation}
%
%\vspace{-5mm}
%
\begin{equation} \label{RamanIn111_E}
\begin{split}
I_{E_{\rm g}} (\hat{\bf e}_{\rm i}, \hat{\bf e}_{\rm s}) \propto \left | \hat{\bf e}_{\rm i}^{\ast} \cdot R^{\prime}(E_{\rm g}^1) \cdot \hat{\bf e}_{\rm s} \right |^{2} + \left | \hat{\bf e}_{\rm i}^{\ast} \cdot R^{\prime}(E_{\rm g}^2) \cdot \hat{\bf e}_{\rm s} \right |^{2},
\end{split}
\tag{A9}
\end{equation}
%
%\vspace{-5mm}
%
\begin{equation} \label{RamanIn111_T}
\begin{split}
I_{T_{\rm 2g}} (\hat{\bf e}_{\rm i}, \hat{\bf e}_{\rm s}) \propto \left | \hat{\bf e}_{\rm i}^{\ast} \cdot R^{\prime}(T_{\rm 2g}^1) \cdot \hat{\bf e}_{\rm s} \right |^{2} + \left | \hat{\bf e}_{\rm i}^{\ast} \cdot R^{\prime}(T_{\rm 2g}^2) \cdot \hat{\bf e}_{\rm s} \right |^{2}\\
 + \left | \hat{\bf e}_{\rm i}^{\ast} \cdot R^{\prime}(T_{\rm 2g}^3) \cdot \hat{\bf e}_{\rm s} \right |^{2}.
\end{split}
\tag{A10}
\end{equation}
In the above equations, $\hat{\bf e}_{\rm i}$ and $\hat{\bf e}_{\rm s}$ represent Jones vectors corresponding to the incident- and scattered-light rays, respectively. 

\bibliographystyle{apsrev4-2}
\bibliography{YIGMagnon}
\end{document}